\documentclass[prl,aps,amssymb,twocolumn]{revtex4}
\usepackage[dvips]{epsfig,color}
\usepackage{graphicx}
\usepackage{amsmath,amsfonts,amsbsy,amssymb}

\begin{document}

\title{Dimensional Hierarchy in Quantum Hall Effects on Fuzzy Spheres}
\author{Kazuki Hasebe$^{1}$ and Yusuke Kimura$^2$}
\affiliation{$^1$Yukawa Institute for Theoretical Physics, Kyoto University,
 Kyoto 606-8502, Japan \\
$^2$Theoretical Physics Laboratory, RIKEN, Wako, Saitama 351-0198, Japan \\
Email: hasebe@yukawa.kyoto-u.ac.jp, kimuray@riken.jp}

\begin{abstract}
We construct higher dimensional quantum Hall systems 
based on fuzzy spheres. 
It is shown that fuzzy spheres are realized  as spheres in colored  monopole backgrounds.
The space noncommutativity is related to higher spins which is 
originated from the internal structure of fuzzy spheres. 
In $2k$-dimensional quantum Hall systems, Laughlin-like wave
function supports fractionally charged
excitations, $q=m^{-\frac{1}{2}k(k+1)}$ (m is odd). 
Topological objects are ($2k-2$)-branes whose statistics are 
determined by the linking number related to the general Hopf map.
Higher dimensional quantum Hall systems exhibit a dimensional hierarchy, where lower dimensional branes condense to make 
higher dimensional incompressible liquid.
\end{abstract}

\maketitle
Zhang and Hu have succeeded to construct a four dimensional generalization of 
 quantum Hall (QH) systems \cite{Zhang2001}.
Their systems have attracted much attention from physicists in various area
\cite{Hu2002,Bernevig2003,Karabali2002,GuowuMeng2003,Kimura2002,Fabinger2002,Sparling0211,Chen0210059,Chen2002,Elvang0209,Kitazawa2002,Chong2003,Bellucci2003PhysLettB,Dolan2003}.
Bernevig et al. also constructed eight dimensional ``spinor'' and ``vector'' QH systems \cite{Bernevig2003}.
They are based on the Hopf map related to the division algebra.
Karabali and Nair have found another way to make higher dimensional  
QH systems based on  $\mathbb{C}P$ manifolds \cite{Karabali2002}.
Their construction originally had nothing to do with the Hopf map, but 
includes the four dimensional and the eight dimensional ``vector'' QH liquid.
$\mathbb{C}P^3$ corresponds to the four dimensional QH liquid 
\cite{Karabali2002}, 
while $\mathbb{C}P^7$ corresponds to the eight dimensional ``vector''
QH liquid \cite{Bernevig2003}. 

One of the intersting features of QH systems is the appearance of the 
 noncommutative geometry.
Coordinates of electrons are noncommutative on the lowest Landau level.
It is well known QH physics in two dimension is governed 
by noncommutative structure \cite{ezawa2003}. 
In Zhang and Hu's work, a four dimensional sphere was not enough to construct a four dimensional
 QH system  
 but an extra isospin space is needed, where 
the need of the isospin space was recognized heuristically.
From the noncommutative geometrical point of view, the need of the extra isospin space
 is naturally understood.
The noncommutative structure applies  symplectic manifolds, which allow 
Poisson structure. 
$\mathbb{C}P$ manifolds are symplectic manifolds, while $S^{2k},k\ge2$ 
are not \cite{NashandSen}.
 Therefore $\mathbb{C}P$ manifolds are enough to  become  
higher dimensional QH systems, while
 $S^{2k},k\ge 2$  need an extra isospin space to incorporate the noncommutative 
structure. 

Therefore it seems reasonable to construct  another kind of higher dimensional QH systems based on ``symplectic spheres'', namely fuzzy spheres, instead of  ``ordinary spheres''. 
The general even dimensional QH systems have been analyzed in 
\cite{GuowuMeng2003}.
However the role of noncommutativity has  not been clarified yet.
In this article,  we construct  higher dimensional QH liquid based 
on  fuzzy spheres.
Our higher dimensional quantum liquid naturally incorporates the
 four dimensional and the eight dimensional ``spinor'' QH liquid.

\vspace{3mm}
Higher dimensional fuzzy spheres $S_F^{2k},~k\ge 2 $ have some characteristic properties 
 which cannot be seen in two dimensional fuzzy spheres.
They are topological equivalent to the  coset $SO(2k+1)/U(k)$
 in a continuum limit, and have  internal fiber spaces which form
 a lower dimensional fuzzy spheres \cite{Ho2002,Kimura2002}.
Therefore, the dimension of the fuzzy $2k$-shere is not $2k$ but 
$k(k+1)$.
$SO(2k+1)/U(k)$ is a symplectic manifold, where 
the noncommutative structure can be incorporated.

The stabilizer group of the $SO(2k)$ spinor is $U(k)$. 
The coset $SO(2k)/U(k)$ is isomorphic to 
 the configuration space of the $SO(2k)$ spinor.
Therefore, it is promising to construct a projection,
\begin{equation}
S_F^{2k}\rightarrow S^{2k},
\end{equation}
with the use of Hopf spinor $\Psi$ on $S^{2k}$.
The explicit projection is established as,
\begin{equation}
 \Psi \rightarrow \frac{X_{a}}{R}=\Psi^{\dagger}\Gamma_{a}\Psi, ~~~~\Psi^{\dagger}\Psi=1,    
\end{equation}
where $\Gamma_a$'s are  Clifford algebra in $(2k+1)$-dimensional space.
$X_a$'s are coordinates on $S^{2k}$, $\sum_{a=1}^{2k+1}X_a^{2}=R^2$; 
the Hopf spinor $\Psi$ is a coordinate on the manifold $U(1)\otimes S_{F}^{2k}$
 as we shall see  below.

The Clifford algebra $\Gamma_{a}$'s  are explicitly represented as, 
\begin{align}
&\Gamma^i = \left(%
\begin{array}{cc}
  0 & i\gamma^i \\
  -i\gamma^i & 0 \\
\end{array}%
\right), \;\;\; i=1,...,2k-1
\nonumber\\
&\Gamma^{2k} = \left(%
\begin{array}{cc}
  0 & 1_{2^{k-1} \times 2^{k-1}} \\
  1_{2^{k-1}\times 2^{k-1}} & 0 \\
\end{array}%
\right),\nonumber\\
&\Gamma^{2k+1} = \left(%
\begin{array}{cc}
  1_{2^{k-1} \times 2^{k-1}} & 0 \\
  0 & -1_{2^{k-1} \times 2^{k-1}} \\
\end{array}%
\right).
\end{align}
The Hopf spinor  $
\Psi=
\begin{pmatrix}
& \Psi_+\\
& \Psi_- \\
\end{pmatrix}$,
can be constructed in the following way,
\begin{align}
&\Psi_+ = \sqrt{\frac{R+X_{2k+1}}{2R}} \Psi^{(2k-2)},\nonumber\\
&\Psi_-=\frac{1}{\sqrt{2R(R+X_{2k+1})}} (X_{2k} -i X_i \gamma^i)\Psi^{(2k-2)},
\label{PsiFuzzySphere}
\end{align}
where $\gamma^i$'s are Clifford algebra in $(2k-1)$-dimensional space.
$\Psi$ is constructed by adding the $2k$ degrees of freedom $X_a,~
a=1,2,\cdots, 2k+1$, which is the coordinate on $S^{2k}$, to $\Psi^{(2k-2)}$.   
``Subcomponent'' $\Psi^{(2k-2)}$ is also comprised of $(2k-4)$ dimensional spinor $\Psi^{(2k-4)}$ just like (\ref{PsiFuzzySphere}). 
Similarly $\Psi^{(2k-4)}$  is comprised of $\Psi^{(2k-6)}$.
Thus, we have such a  sequence, 
\begin{equation}
\Psi\rightarrow \Psi^{(2k-2)}\rightarrow \Psi^{(2k-4)}\rightarrow\cdots \rightarrow\Psi^{(2)}.
\end{equation}
The normalization $\Psi^{\dagger}\Psi\!\!=\!\!1$  
stems from a subcomponent normalization $\Psi^{(2k-2)\dagger}\Psi^{(2k-2)}\!\!=\!\!1$, eventually from  the minimal subcomponent normalization $\Psi^{(2)\dagger}\Psi^{(2)}\!\!=\!\!1$. 
Since the Hopf spinor $\Psi$ is constructed by adding the
 degrees of freedom of subcomponent spheres to $\Psi^{(2)}$, it  
 has such degrees of freedom; $2k+(2k-2)+\cdots +2+1\!=\!k(k+1)+1$.
 It is equal to  $S_F^{2k}\otimes U(1)$ degrees of freedom as it should be.  

The holonomy group of the base manifold $S^{2k}$ is $SO(2k)$.
Then, the spinor which lives on  $S^{2k}$ is described as a section of the 
  $Spin(2k)$ bundle on $S^{2k}$.
Berry connection $A$, which is obtained from the Hopf spinor as $
{\Psi}^{\dagger}d\Psi={\Psi^{(2k-2)}}^{\dagger}\cdot i A_{a}dX_a \cdot\Psi^{(2k-2)}$,
is actually  a $Spin(2k)$ connection,
\begin{equation}
A_{\mu}=-\frac{1}{R(R+X_{2k+1})}\Sigma^{+}_{\mu\nu}X_{\nu},~~~~A_{2k+1}=0, 
\label{Monogauge}
\end{equation}
where $\Sigma_{\mu\nu}$'s are $Spin(2k)$ generators, 
\begin{equation}
\Sigma_{\mu\nu}=-i\frac{1}{4}[\Gamma_{\mu},\Gamma_{\nu}]=
\begin{pmatrix}
  \Sigma_{\mu\nu}^{+} & 0 \\
  0 & \Sigma_{\mu\nu}^{-} 
\end{pmatrix},\; \mu,\nu = 1,..,2k, 
\label{sigmapm}
\end{equation}
in detail, $\Sigma_{\mu\nu}^{\pm}=\{\Sigma^{\pm}_{ij},\Sigma^{\pm}_{2k,i}\}
=\{-i\frac{1}{4}[\gamma_i,\gamma_j],\mp\frac{1}{2}\gamma_i\}.$
We locate an $Spin(2k)$ colored monopole at the center of $S^{2k}$ to be
 compatible  with  the  geometrical $SO(2k)$ holonomy,  which 
generates the $Spin(2k)$ gauge connection on $S^{2k}$.
We identify this monopole  gauge connection with the spinor geometrical connection.
Consequently, the magnitude of the spin $I$,  in other words $SO(2k+1)$ spinor 
irreducible representation index, is related to the monopole charge $g$. 

The colored 
monopole  has been studied in \cite{Horvath1978,Yang1978}, 
which is transformed to a $Spin(2k)$ colored instanton 
on $2k$-dimensional Euclidean space 
 by a
stereographic projection.
Such an  instanton configuration satisfies a higher dimensional
 dual equation  and its
 stability is guaranteed by the homotopy mapping,
 $\pi_{2k-1}(SO(2k))=Z$ \cite{Saclioglu1986}.
The topological number  is described by
the $k$-th Chern number $I$, physically which  corresponds to  the colored monopole charge $g$ as $~
g=\frac{1}{2}I,~I\in \text{integer}$.
For instance,
the gauge connection (\ref{Monogauge}) yields a unit element of the 
Chern number, or the monopole charge $1/2$.
The monopole whose charge is $g={I}/{2}$ is easily obtained 
from (\ref{Monogauge}) by using the $(0,\cdots,0,I)$ representation of $Spin(2k+1)$. 
Due to the gauge and geometrical holonomy identification, the monopole 
charge is equivalent to the magnitude of the  particle spin.
Similarly, the $Spin(2k)$ colored monopole generators can be regarded as the 
particle spins.   

We construct higher dimensional QH systems based on this setup. 
The  particle Hamiltonian   is 
\begin{equation}
H=\frac{\hbar^{2}}{2MR^2}\sum_{a<b}\Lambda^2_{ab},
\label{particleHamil}
\end{equation}
where $\Lambda_{ab}$ is the covariant particle angular momentum, 
$\Lambda_{ab}=-i(X_a D_b-X_b D_a)$.
$D_a$ is the covariant momentum,
$D_{a}=\partial_a+iA_a$. 
The algebraic relation of the covariant particle angular momentum 
$\Lambda_{ab}$ is 
\begin{align}
\!\!\!\!\!\!&[\Lambda_{ab},\Lambda_{cd}]= i [\delta_{ac} \Lambda_{bd} + \delta_{bd}\Lambda_{ac} -
\delta_{bc} \Lambda_{ad} - \delta_{ad} \Lambda_{bc}]\nonumber\\
\!\!\!\!\!\!&-\!\!i [X_{a}X_{c} F_{bd} + X_{b}X_{d}F_{ac} -
X_{b}X_{c}F_{ad} - X_{a}X_{d} F_{bc}].
\label{lambdaalgebra}
\end{align}
$\Lambda_{ab}$'s  do not satisfy the $SO(2k+1)$ algebra, due to 
the existence of the  monopole field strength $F_{ab}$.

Monopole angular momentum is $R^2 F_{ab}$, whose explicit form is 
$R^2 F_{\mu\nu}
=-(X_{\mu}A_{\nu}-X_{\nu}A_{\mu}-{\Sigma}^+_{\mu\nu}),
~R^2 F_{\mu 2k+1}=(R+X_{2k+1})A_{\mu}$.
They satisfy same algebra as in (\ref{lambdaalgebra})
 with the replacement $\Lambda_{ab}\rightarrow R^2 F_{ab}$.
The magnitude of the monopole angular momentum is equal to 
the $Spin(2k)$ quadratic Casimir,
$R^4 F_{ab}^2=\Sigma^{+2}_{\mu\nu}$.
The particle angular momentum $\Lambda_{ab}$ is parallel to the
 tangent space of the base manifold $S^{2k}$, while 
the monopole angular momentum $R^2 F_{ab}$ is orthogonal to it.
Therefore they are orthogonal in each other,
$\Lambda_{ab}\cdot R^2 F_{ab}=R^2 F_{ab}\cdot \Lambda_{ab}=0.$

The total angular momentum is the sum of the particle and the monopole angular momentum,$
L_{ab} =\Lambda_{ab}+R^2 F_{ab}$,
 in detail, $
L_{\mu \nu} = L_{\mu \nu}^{(0)} +  \Sigma^+_{\mu \nu},~
L_{\mu 2k+1} = L_{\mu 2k+1}^{(0)} + RA_\mu + X_{\mu}A_{2k+1}$,
which satisfy the $SO(2k+1)$ algebra.

The particle Hamiltonian (\ref{particleHamil}) commutes with the 
 total angular momentum $L_{ab}$ due to the  $SO(2k+1)$ symmetry in the system.
It is rewritten as $
H=\frac{1}{2MR^2}\sum_{a<b}(L^2_{ab}-R^4 F^2_{ab}),$
where the  orthogonality of 
the particle and the  monopole angular momentum was used.
The eigenenergy of $Spin(2k+1)$  representation $(n,0,\cdots,0,I)$  is 
\begin{equation}
E(n,I)=\frac{\hbar^2}{2MR^2}[n^2+n(I+2k-1)+\frac{1}{2}Ik],
\label{LLenergy}
\end{equation}
where $n,I=0,1,2,\cdots$.
$n$ indicates Landau levels, while  
$I$  determines the degeneracies in each Landau level.

The lowest Landau level (LLL) corresponds to  $n=0$.
The energy and the degeneracy  are given as 
\begin{align}
&E_{LLL}=\frac{\hbar^2}{2M}\frac{I}{2R^2}k\label{LLLenergy},\\
&d(I)\!\!=\!\!\frac{(I+2k-1)!!}{(2k-1)!!(I-1)!!}\!\!\prod _{l=1}^{k-1}\!\!\frac{(I+2l)! l!}{(I+l)!(2l)!}\!\approx \!I^{\frac{1}{2}k(k+1)}.\label{LLLdegene}
\end{align}
The increase of the monopole charge $I/2$ induces the 
 number of  Dirac flux quanta or   states as in (\ref{LLLdegene}).
The  eigenstates in the LLL are given as 
the $SO(2k+1)$ spinors $\{\Phi_{\alpha}\}\in (0,\cdots,0,I)$,
 which is constructed from the 
$I$-fold symmetric products of the Hopf spinor.
It is because the Hopf spinor 
is the $SO(2k+1)$ fundamental spinor $(0,\cdots,0,1)$, which can be explicitly
 proven  as $
L_{ab}^2\Psi_{\pm}=\Sigma_{ab}^{\pm 2} \Psi_{\pm}$ with the use of 
(\ref{PsiFuzzySphere}),(\ref{sigmapm}).

Slater antisymmetric  wavefunction for many particles is given as 
\begin{equation}
\Phi_{Slater}=\sum_{\alpha_1,...,\alpha_{N}}\epsilon_{\alpha_1,...,\alpha_N}
\Phi_{\alpha_1}(x_1)....\Phi_{\alpha_N}(x_N),
\label{Slater}
\end{equation}
where $N$ is the number of particles. 
The Laughlin-like wave function is  obtained 
from (\ref{Slater}) as 
\begin{equation}
\Phi_{Laughlin}=\Phi_{Slater}^m.
\label{Laughlin}
\end{equation}
The power $m$ should be taken an odd integer to keep the antisymmetricity of  electrons.

The Slater function (\ref{Slater}) has a property
 of  incompressibility.
It can be  observed from the radial distribution function,
\begin{align}
g(x,x')&=\int dx_3 \cdots dx_{N}
|\Phi_{Slater}(x,x',x_3,\cdots,x_N)|^2\nonumber\\
&\approx 1-\prod_{l=1}^{k}\exp\biggl(-\frac{1}{2\ell_B^2}{(X^{\mu}_{(2l)}-X^{'\mu}_{(2l)})^2\biggr)},
\end{align}
where $x=(X_{(2)},X_{(4)},\cdots,X_{(2k)})$;
$X_{(2l)}$ represents the coordinates on the subcomponent sphere $S^{2l}$.
In the second line, we have expanded the equation around the ``north pole''; $X^{3}_{(2)},X^{5}_{(4)},\cdots,X_{(2k)}^{2k+1}\!\approx\! R$.
The radial distribution  function converges exponentially to unity due to the 
strong suppression of density fluctuation, which is a typical property of the incompressible liquid \cite{Girvin84PhysRevB}.
Intriguingly, our liquid exhibits the incompressibily not only in the 
$2k$-dimensional base space but also in each of the subcomponent 
lower dimensional spaces.
This suggests that
 the higher dimensional QH liquid consists of
lower dimensional incompressible liquid.
We will revisit this later.

The thermodynamic limit naively corresponds to  
$I,R\rightarrow\infty$.
To make the energy (\ref{LLenergy}) finite, we take such a limit keeping the magnetic length $\ell_B=R\sqrt{\frac{2}{I}}$ finite.
The filling factor $\nu$ is
\begin{equation}
\nu=\frac{N}{d(mI)}\approx m^{-\frac{1}{2}k(k+1)},
\label{highernu}
\end{equation} 
where  $N=d(I)$.
 The density behaves as $
\rho={N}/{V}\approx {d(I)}/{R^{k(k+1)}}\approx \ell^{-k(k+1)}_B$,
which is finite in the thermodynamic limit as it should be.
Due to (\ref{highernu}), 
Laughlin-like wave function   supports excitations,  whose charge is fractional,
$q=m^{-\frac{1}{2}k(k+1)}$. 

Around the ``north pole'', the higher dimensional QH 
Lagrangean is reduced to 
\begin{equation}
L=\frac{dX^{\mu}}{dt}A_{\mu} -V,
\end{equation}
where we have added a confining potential term $V$.
The gauge field is 
$A_{\mu}=-\frac{1}{\ell_B^2}\Sigma_{\mu\nu}^{+}X_{\nu}$.
$\Sigma_{\mu\nu}^+$'s can be  expanded by a linear combination of the 
$Spin(2k)$ generators, $t^a, a=1,2,\cdots,2k^2-3k+1$ as  
$\Sigma_{\mu\nu}^{+}=(2k-1)\eta_{\mu\nu}^a t^a$.
 $\eta_{\mu\nu}^a$ is an expansion coefficient, namely a generalized t'Hooft symbol. $(2k-1)$ is  for convention.
  This expansion is always possible because it is   just a recombination of basis.
For example, at $k=2$, $t^a$'s are $SU(2)$
 spin generators, at $k=3$, $SU(4)$ spin generators.

Canonical momentum reads
$P_{\mu}=-\frac{1}{\ell_B^2}\Sigma^{+}_{\mu\nu}X_{\nu}$.
The coordinate and the momentum are orthogonal, due to the antisymmetry
 of the $SO(2k)$ generator, which physically corresponds to 
  a higher dimensional cyclotron motion.

By imposing the canonical quantization condition, we obtain 
noncommutative algebra in higher dimensional QH systems,
\begin{equation}
[X_{\mu},X_{\nu}]=-2i\ell_B^2\eta_{\mu\nu}^a t^a,
\label{NCandspin}
\end{equation}
which is consistent with the algebra obtained from the higher dimensional fuzzy spheres \cite{Ho2002}.
Since  $\{t^a\}$ construct $S^{2k-2}_F$ algebra,
 it is found that the noncommutativity of coordinates is related 
to the lower dimensional fuzzy sphere $S_F^{2k-2}$ \cite{Kimura2002}.
It is natural to regard this lower dimensional fuzzy sphere as an internal space.
Therefore, electrons in the higher dimensional QH systems have effective 
 higher spins.
Thus, two important quantities, the space noncommutativity and the $Spin(2k)$ group spin, are related by the ``fundamental  length'' $\ell_B$ as in (\ref{NCandspin}).\\
The equation of motion reads
\begin{equation}
\frac{d}{dt}X_{\mu} =-i[X_{\mu},H]=-2\ell^2_B
\eta_{\mu\nu}^{a}t^{a}\frac{\partial V}{\partial X_{\nu}}.
\label{spinHallcurrents}
\end{equation}
The Hall current $\frac{d}{dt}X_{\mu}$ along a spin direction is determined by this equation, which  is orthogonal to the ``electric field'' 
$-\frac{\partial V}{\partial X_{\mu}}$ as expected.

The subcomponent spheres $S^{2k-2}$ on the base manifold $S^{2k}$ are regarded as 
$(2k-2)$-branes. 
$(2k-2)$-branes  move on the $2k$-dimensional
 base space. Then, the  total dimension of 
the space-time  where $(2k-2)$-brane exists is to be considered as  $(4k-1)$.
The statistics of $(2k-2)$-branes is
 determined by the general Hopf map $\pi_{4k-1}(S^{2k})=\mathbb{Z},~
k\in \mathbb{N}$ \cite{Wu1988PhysLettB}.
(Strictly speaking, there sometimes appears an extra discrete group 
from the torsion part which we have omitted in the RHS.
  For instance, $\pi_7(S^4)=\mathbb{Z}\oplus\mathbb{Z}_{12}$. However it is  not relevant in the following discussion.) 
The topological number 
represents the  linking number of the $(2k-2)$-branes 
in the $(4k-1)$-dimensional space-time \cite{Bott}. 
They are higher dimensional generalizations of the  
 anyonic quasiparticles in two dimensional QH systems, whose statistics  is determined by the braid group in 3-dimension 
\cite{Wilczek83PhysRevLett}.

\begin{figure}[tbph]
\includegraphics*[width=65mm]{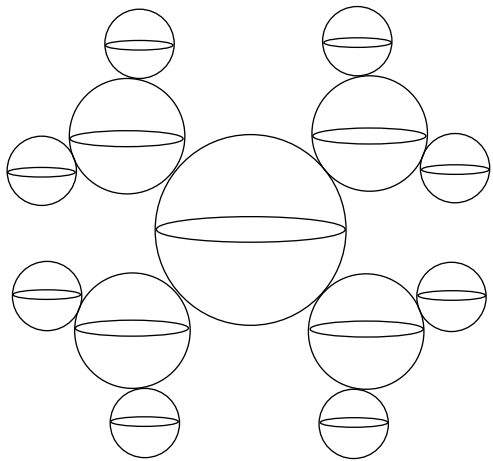}
\caption{ Lower dimensional branes iteratively condense
 to make higher dimensional incompressible liquid. }
\label{Fuzzysphere}
\vspace{-3mm}
\end{figure}

Fractional QH states  respect the Haldane-Halperin hierarchy \cite{Haldane1983}.
In the dual picture, this hierarchy is understood as a formation of 
new Laughlin states by quasiparticles or 0-branes \cite{Lee1991}. 
Here, we see that QH systems also exhibit another kind of hierarchy, namely a dimensional
 hierarchy. 
For instance, 0-branes which live
 on $S^{2}$ contruct the two dimensional QH liquid. 
In general, on the subsphere $S^{2l}$, each $(2l-2)$-brane 
occupies an area of $\ell_B^{2l}$.
The number of the $(2l-2)$-brane on $S^{2l}$
  is ${R^{2l}}/{\ell^{2l}_B}\approx I^l$.
These $I^l$ $(2l-2)$-branes condense to make new 
$2l$-dimensional incompressible liquid. 
Iteratively, the lower dimensional incompressible liquid condenses to 
make higher dimensional incompressible liquid.
As a result, $2k$-dimensional QH liquid is made of $I^{\frac{1}{2}(k-1)k}$ 
2-dimensional incompressible liquid.
The higher dimensional filling factor (\ref{highernu}) is naturally  interpreted
 in this context.
The filling factor on the subcomponent sphere $S^{2l}$  is 
$I^l/(mI)^l=m^{-l}.$
The filling factor $\nu$ in the total $S_F^{2k}$ space is obtained by 
a product of the each filling factor in the subcomponent spheres,
\begin{equation}
\nu=\frac{1}{m}\cdot \frac{1}{m^2}\cdot \frac{1}{m^3}\cdots \frac{1}{m^{k}}=
m^{-\frac{1}{2}k(k+1)}.
\end{equation}
Thus, the higher dimensional filling factor represents the dimensional 
condensation of lower dimensional QH liquid [Fig.\ref{Fuzzysphere}].
This dimensional hierarchy implies an intimate connection 
between the QH systems and  the matrix models.
In the matrix theory picture, 
the higher dimensional D-brane is comprised   of  lower dimensional 
D-branes \cite{Banks1997,Ishibashi1997}.
The general filling factor is given by the combination of the dimensional and the Haldane-Halperin hierarchy, 
\begin{align}
&\nu=
\frac{1}{m\pm \frac{1}{2p_1\pm\frac{1}{2p_2\pm\cdots\frac{1}{2p_f}}}}
\cdot 
\frac{1}{m^2\pm \frac{1}{(2p_1)^2\pm\frac{1}{(2p_2)^2\pm\cdots\frac{1}{(2p_f)^2}}}}
\nonumber\\
&~~~~\cdots
\frac{1}{m^k\pm \frac{1}{(2p_1)^k\pm\frac{1}{(2p_2)^k\pm\cdots\frac{1}{(2p_f)^k}}}},
\end{align}
where in  each space dimension, 
 branes exhibit generalized a Haldane-Halperin 
hierarchy.

The edge of the $2k$-dimensional QH liquid  is $S^{2k-1}$, whose symmetry group is 
 $SO(2k)$.
As in the four dimensional case \cite{Hu2002}, there will also appear various $Spin(2k)$ higher spin particles 
  as dipoles due to the nontrivial $Spin(2k)$
 composition rule.
The edge action is described by the Wess-Zumino-Witten
 model \cite{Karabali030}.
 Therefore, our edge states will arise as excitations in the $Spin(2k)$ Wess-Zumino-Witten model.
We will report the detailed analysis of this model in the forth coming paper.

\vspace{3mm}
The authors would like to acknowledge  the
 theoretical physics laboratory in RIKEN for the helpful atomosphere.
They also express their gratitude to S.Iso for fruitful discussions and comments.
K.H. is very glad to T.Fukuda, H.Kajiura, B-H.Lee, H-C.Lee, K.Moon, S.Park, C.Rim, K.Shizuya, S.Sugimoto, H-S.Yang  and  J.Yeo for useful discussions.  
He also thanks  the string theory group members in Sogang University 
for their hospitality. 
K.H. was supported by grant No. R01-1999-00018 from the
Interdisciplinary Research Program of the KOSEF, the special grant
of Sogang University in 2002 and JSPS fellowship.

\vspace{-5mm}

\end{document}